\documentclass[10pt,pra,twocolumn,preprintnumbers,aps]{revtex4}
\usepackage{amssymb}
\usepackage{amsfonts}
\usepackage{amssymb}
\usepackage{amsmath}
\usepackage[dvips]{graphicx}
\usepackage{color}
\usepackage{ulem}

\begin{document}

\title{Experimental preparation and manipulation of squeezed cat states via an all-optical in-line squeezer}
\author{Meihong Wang$^{\ddagger 1,2}$}
\author{Miao Zhang$^{\ddagger 1,2}$}
\author{Zhongzhong Qin$^{1,2}$}
\email{zzqin@sxu.edu.cn}
\author{Qiang Zhang$^{1,2}$}
\author{Li Zeng$^{1,2}$}
\author{Xiaolong Su$^{1,2}$}
\email{suxl@sxu.edu.cn}
\author{Changde Xie$^{1,2}$}
\author{Kunchi Peng$^{1,2}$}

\affiliation{$^1$State Key Laboratory of Quantum Optics and Quantum Optics Devices,
Institute of Opto-Electronics, Shanxi University, Taiyuan 030006, People's
Republic of China\\
$^2$Collaborative Innovation Center of Extreme Optics, Shanxi University,
Taiyuan, Shanxi 030006, People's Republic of China\\
}

\begin{abstract}
{The squeezed cat state, an essential quantum resource, can be used for quantum error correction and slowing
decoherence of the optical cat state. However, preparing a squeezed cat state with high generation rate, and effectively manipulating it, remain challenging. In this work, a high-performance all-optical in-line squeezer is developed to prepare a squeezed cat state and manipulate the phase of the quadrature squeezing.  
This scheme has the advantages that the phase of the quadrature squeezing of the squeezed cat state can be manipulated by changing the working condition of the squeezer, and that a higher generation rate can be achieved via the deterministic squeezing operation of the in-line squeezer. The generation rate of squeezed cat states reaches 2 kHz, the same as that of the initial cat state. The all-optical in-line squeezer proposed here removes the requirements of electro-optic and opto-electric conversions necessary for an off-line squeezer, thus enabling high-bandwidth squeezing operations on non-Gaussian states. These results provide an efficient method to prepare and manipulate optical squeezed cat states, which makes a step closer to their applications in all-optical quantum information processing.}
\end{abstract}

\maketitle

\section{Introduction}

The Schr\"{o}dinger cat state plays an important role in exploring the boundary between quantum and classical physics \cite{Schrodinger1935,HarocheRMP2013,Arndt2014}, quantum information science \cite{RalphPRA2003,JeongPRA2002,LundPRL2008,QEC2022,vanEnkPRA2001,Hacker2019}, and quantum metrology \cite{JooPRL2011,Gilchrist2004}.
 It has been prepared in diverse systems, such as cavity quantum electrodynamics \cite{Raimond1997}, ion traps \cite{Monroe1996}, superconducting quantum circuits \cite{Vlastakis2013,chaosong2019}, and Rydberg atom arrays~\cite{omran2019}. Free-propagating optical Schr\"{o}dinger cat states have attracted much attention due to their weak interaction with the environment, which is beneficial for quantum information processing \cite{RalphPRA2003,JeongPRA2002,LundPRL2008,QEC2022,vanEnkPRA2001,Hacker2019}. An optical cat state is defined as a superposition of two coherent states  
$\left\vert \alpha \right\rangle$ and $\left\vert -\alpha \right\rangle$ with opposite phases and same mean photon number $\left\vert \alpha\right\vert ^{2}$. The overlap between the two coherent-state components is $\left\vert \left\langle \alpha\right\vert -\alpha \rangle \right\vert ^{2}=e^{-4\left\vert \alpha\right\vert ^{2}}$, which decreases exponentially with the increase of $\left\vert \alpha\right\vert $. 

~~To prepare an optical cat state, a frequently used method is subtracting photons from a squeezed vacuum state \cite{Dakna1997,Lund2004,Ourjoumtsev2006,Neergaard2006,LeeScience2011,ZhangM2021,Takahashi2008,Gerrits2010,Laghaout2013}. 
As a typical non-Gaussian state of light, optical cat states have been experimentally prepared by subtracting one photon~\cite{Ourjoumtsev2006,Neergaard2006,LeeScience2011,ZhangM2021}, two photons~\cite{Takahashi2008} and three photons~\cite{Gerrits2010} from squeezed vacuum states. Since a squeezed vacuum state is a continuous-variable quantum resource and the photon detection is a typical discrete-variable technique, the preparation of an optical cat state involves hybrid quantum information processing techniques \cite{Loock2011,Hybrid,Do2021}. The  prepared optical cat states have been applied in quantum teleportation of cat states~\cite{LeeScience2011}, tele-amplification~\cite{Neergaard2013}, preparation of hybrid entangled states~\cite{Jeong2014,Morin2014,Ulanov2017,Sychev2018}, 
and the Hadamard gate~\cite{Tipsmark2011}.

~~Besides optical cat states, squeezed cat (SC) states have also been identified as valuable resources for fault-tolerant quantum information processing. For example, it has been shown that SC states have advantages in quantum error correction~\cite{QEC2022} and slowing 
decoherence of optical cat states~\cite{Serafini2004,SQDPRL2018}. In quantum error correction, the SC code allows to correct the errors caused by photon loss, while at the same time improving the protection against dephasing~\cite{QEC2022}. To slow decoherence of optical cat states in quantum communication \cite{SQDPRL2018}, an optical cat state should be squeezed along the superposition direction before the transmission, and then squeezed orthogonal to the superposition direction after the transmission. Up to now, only the SC state squeezed along the superposition direction of coherent states has been prepared~\cite{SQDPRL2018,HuangK2015,Ourjoumtsev2007,Etesse2015,Lvovsky2017}. How to prepare a SC state squeezed orthogonal to the superposition direction, as well as 
effectively manipulate the phase of the quadrature squeezing of SC states, remain challenging. 

~~Another challenge encountered in the current applications of optical SC states is the low generation rate. In 
previous experiments of preparing optical SC states, either a two-mode squeezed state followed by photon detection \cite{SQDPRL2018,HuangK2015}, or a two-photon Fock state followed by homodyne heralding \cite{Ourjoumtsev2007}, was used. In most of these experiments, the generation rates of SC states are in the range of several Hz to 200 Hz~\cite{SQDPRL2018,HuangK2015,Ourjoumtsev2007,Etesse2015}. Thus, it is urgent to develop a more efficient method to improve the generation rate of optical SC states.
\begin{figure*}[htbp]
\begin{center}  \includegraphics[width=0.65\linewidth]{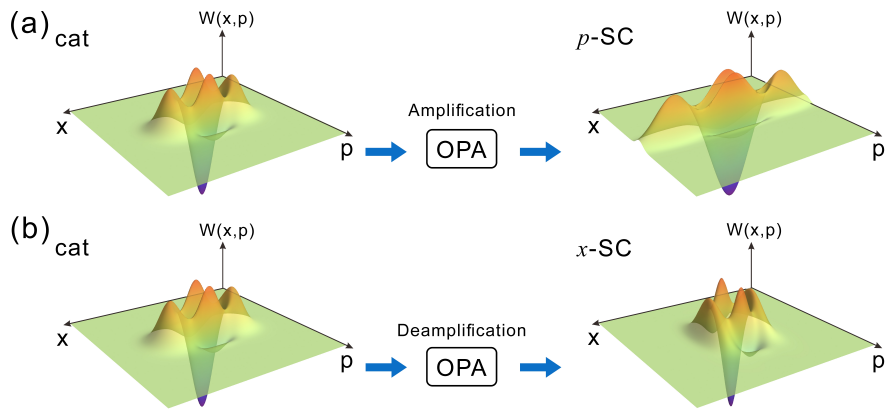}
 \end{center}
  \caption{Principle for preparing optical SC states. a) The principle of preparing a $p$-SC state by an OPA working at the condition of amplification. 
b) The principle of preparing an $x$-SC state by an OPA working at the condition of deamplification. 
OPA, optical parametric amplifier.}
  \label{fig1}
\end{figure*}

~~It has been shown that an optical squeezer can be used to effectively and deterministically manipulate quantum states~\cite{Miwa2014,Zhang2008,Yan2012,Sun2022}. There are two types of optical squeezers, namely the in-line squeezer and the off-line squeezer, depending on whether the input state is coupled into the squeezer directly or not. The measurement-based off-line squeezer has been used to demonstrate the conversion between a single-photon state and an optical cat state~\cite{Miwa2014}. The all-optical measurement-free in-line squeezer has been used to enhance the squeezing and entanglement of Gaussian states~\cite{Zhang2008,Yan2012}. Comparing with the off-line squeezer \cite{Miwa2014}, the all-optical in-line squeezer  
does not require electro-optic and opto-electric conversions, thus lifting the bandwidth limitation imposed by the homodyne detectors and electro-optic modulators. However, it remains a challenge to prepare a SC state with the deterministic in-line squeezer.

~~Here, we develop a scheme to prepare an optical SC state with a high generation rate and to manipulate the phase of the quadrature squeezing via a high-performance all-optical in-line squeezer. By optimizing the configuration and parameters of the in-line squeezer, we achieve a highly efficient, broadband squeezing operation on the input optical cat state, without losing its non-classicality. The phase of the quadrature squeezing of optical SC states is manipulated by changing the working condition of the in-line squeezer. Thanks to the deterministic squeezing operation of our in-line squeezer, the generation rate of optical SC states only depends on that of the initial optical cat state, which is about $2$ kHz. The prepared SC state in our experiment matches rubidium transition line, promising applications in quantum memories based on atomic ensembles \cite{NPOQM2009,YanZHNC2017,Wei2022}. Our results provide an all-optical method to prepare and manipulate SC states. This constitutes a crucial step towards all-optical quantum information processing based on SC states.
\begin{figure*}[htbp]
\begin{center}  \includegraphics[width=0.8\linewidth]{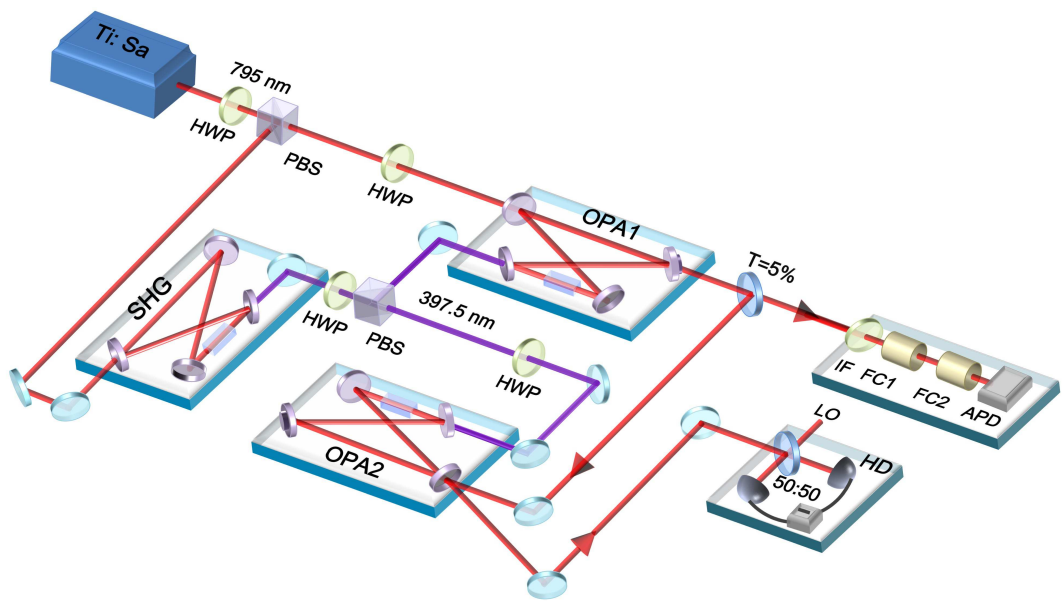}
 \end{center}
  \caption{Experimental setup of preparing and manipulating optical SC states. A photon click in APD heralds photon subtraction from the squeezed vacuum state, which is realized by the beam-splitter with a transmittance of $T = 5\%$. SHG, second harmonic generator; HWP, half-wave plate; PBS, polarization beam splitter; OPA, optical parametric amplifier; IF, interference filter with $0.4$ nm bandwidth; FC, filter cavity;  LO, local oscillator; APD, avalanche photodiode; HD, homodyne detector.}
  \label{fig2}
\end{figure*}

\section{The principle}

To prepare a SC state, a squeezing operation needs to be performed on a cat state. Here, an optical parametric amplifier (OPA) 
is used as an in-line squeezer to implement the squeezing operation, as shown in Figure \ref{fig1}a and \ref{fig1}b. The Wigner function of an optical odd cat state, whose superposition direction is along the $x$ quadrature, is given by 
\begin{eqnarray}
W_{cat} &=&\frac{1}{\pi N_{-}}\big[e^{-(x-\sqrt{2}\alpha )^{2}-p^{2}}+e^{-(x+\sqrt{2}\alpha
)^{2}-p^{2}} \notag \\
& &\ \ \ \ \ \ \ \ \ \ \ -2e^{-x^{2}-p^{2}}cos(2\sqrt{2}p\alpha )\big] 
\end{eqnarray}
where $\alpha$ is the amplitude of the optical cat state, $N_{-}=2-2e^{-2\alpha ^{2}}$ is the normalization factor, $x$ and $p$ are the phase-space amplitude and phase quadratures (position and momentum parameters), respectively. 

~~A $p$-SC state, i.e., squeezed along the $p$ quadrature (orthogonal to the superposition direction by controlling the phase of the quadrature squeezing to $\pi/2$), is prepared when 
the OPA works under the condition of amplification. The Wigner function of the $p$-SC state is given by \cite{Filip2001} 
\begin{eqnarray}
W_{psc} &=&\frac{1}{\pi N_{-}}\big[e^{-\frac{(x-\sqrt{2}e^{r}\alpha )^{2}}{e^{2r}}-\frac{p^{2}}{%
e^{-2r}}}+e^{-\frac{(x+\sqrt{2}e^{r}\alpha )^{2}}{e^{2r}}-\frac{p^{2}}{%
e^{-2r}}} \notag \\
& &\ \ \ \ \ \ \ \ \ \ \ -2e^{-\frac{x^{2}}{e^{2r}}-\frac{p^{2}}{e^{-2r}}}cos(2\sqrt{2}pe^{r}\alpha
)\big]
\end{eqnarray}%
where $r$ is the squeezing parameter of the SC state. Since an in-line squeezer is applied to prepare SC states, the squeezing parameter of the SC state can be controlled by changing the gain of the OPA. 

~~On the other hand, an $x$-SC state, i.e., squeezed along the $x$ quadrature (along the superposition direction by controlling the phase of the quadrature squeezing to $0$), can be prepared when the OPA works under the condition of deamplification. The Wigner function of the $x$-SC state is \cite{Filip2001}
\begin{eqnarray}
W_{xsc} &=&\frac{1}{\pi N_{-}}\big[e^{-\frac{\left( x-\sqrt{2}e^{-r}\alpha \right) ^{2}}{e^{-2r}}-%
\frac{p^{2}}{e^{2r}}}+e^{-\frac{\left( x+\sqrt{2}e^{-r}\alpha \right) ^{2}}{%
e^{-2r}}-\frac{p^{2}}{e^{2r}}} \notag \\
& &\ \ \ \ \ \ \ \ \ \ \ -2e^{^{-\frac{x^{2}}{e^{-2r}}-\frac{p^{2}}{e^{2r}}}}cos\left( 2\sqrt{2}%
pe^{-r}\alpha \right) \big]
\end{eqnarray}
From the theoretical Wigner functions of the $x$-SC and $p$-SC states in Figure \ref{fig1}, it is obvious that the cat state is squeezed along the amplitude and phase quadratures, respectively.

\section{Experimental results}

\begin{figure*}[htbp]
\begin{center}
\includegraphics[width=0.9\linewidth]{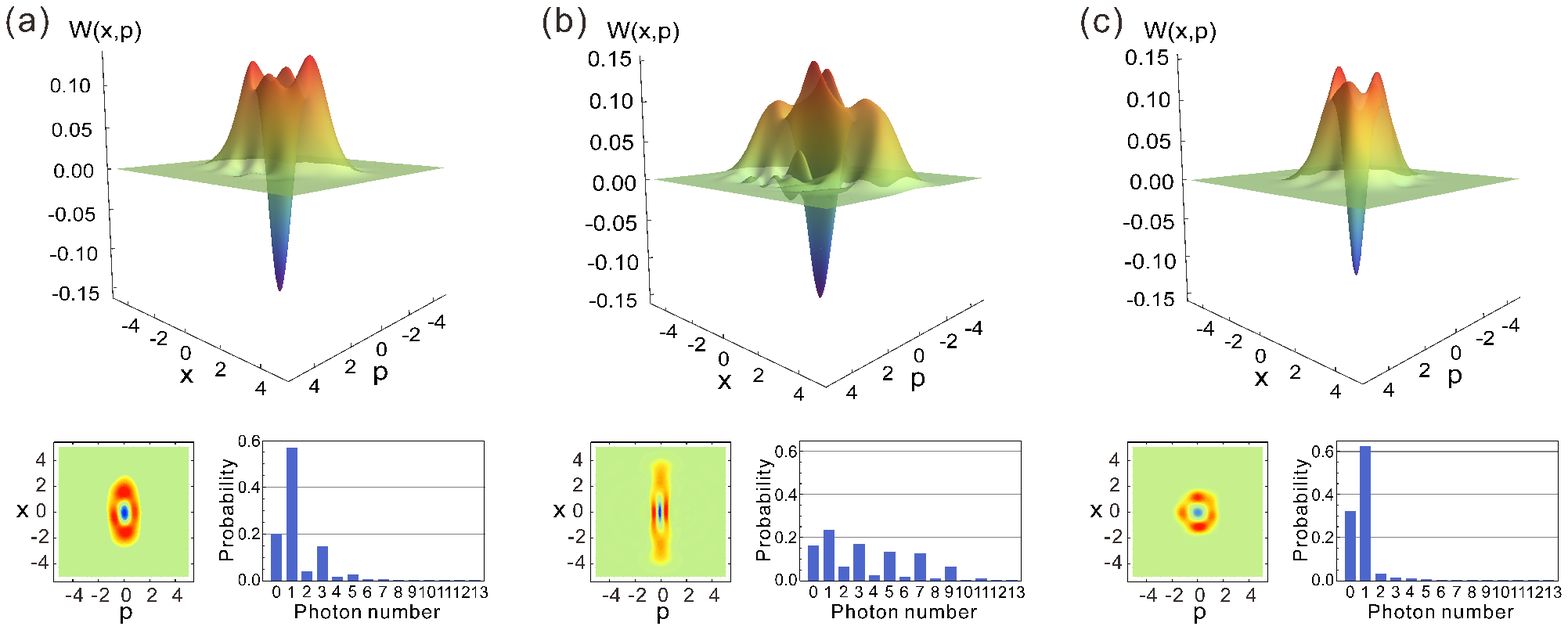}
\end{center}
\caption{Results of cat state and SC states. a) Results of the cat state. b) Results of the $p$-SC state with $40$ mW pump power for OPA2. c) Results of the $x$-SC state with $20$ mW pump power for OPA2. Top: 
experimentally reconstructed
Wigner functions of cat state with $\alpha=1.06$, the $p$-SC state with $\alpha=1.40$ and $r=0.30$, and the $x$-SC state with $\alpha=0.99$ and $r=0.29$,                 
respectively. Bottom: projections of experimentally reconstructed Wigner functions and corresponding photon-number distributions. All results are corrected for a 80\% detection efficiency.}
\label{fig3}
\end{figure*}

As shown in Figure \ref{fig2}, part of the laser beam from a continuous wave Ti: Sapphire laser operated at $795$ nm is sent to a second harmonic generator (SHG) to generate the pump beams at $397.5$ nm for two OPAs, and the rest of it is used as the seed beam of OPA1 and the local oscillator (LO) of homodyne detector. By subtracting a photon from a nearly pure squeezed vacuum state with $-3$ dB squeezing produced by OPA1, the optical cat state is conditionally prepared. Then it is seeded into an in-line squeezer OPA2 for preparing and manipulating the SC state. By controlling the working condition of OPA2 to the condition of amplification or deamplification of the seed beam, i.e., 
locking the phase difference between the seed and pump beams of OPA2 to $0$ or $\pi$, the cat state is squeezed along the phase or amplitude quadrature, respectively. Then, the prepared cat state and SC states are measured by the homodyne detector, respectively. 
The Wigner functions and density matrices of the cat state and SC states are reconstructed by homodyne tomography \cite{Lvovsky2009} (see Section S1, Supporting Information for more details). 

~~In our experiment, two technical challenges have to be solved. One is that the bandwidth of the in-line squeezer should be broad enough, otherwise the information of the cat state will be lost \cite{Miwa2014}. In the experiment, the bandwidth of OPA2 is about $15$ MHz, 
slightly broader than that of OPA1 ($13$ MHz), so that all the information of the cat state can be amplified or deamplified by OPA2. The other technical challenge is that the loss introduced by the in-line squeezer should be as low as possible to avoid the decrease of the non-classicality of the cat state. In our experiment, by optimizing the transmissivity of input (output) coupler of OPA2 $\left( T=14.7\%\right)$, the transmission efficiency of OPA2 is improved to $91\%$, i.e., the loss of OPA2 is reduced to $9\%$.

~~The results of the cat state, the $p$-SC and $x$-SC states, are shown in Figure \ref{fig3}a, \ref{fig3}b, and \ref{fig3}c, respectively. As shown in Figure \ref{fig3}a, the Wigner function of the cat state shows two positive Gaussians of $\left\vert \alpha \right\rangle $ and $\left\vert -\alpha\right\rangle $, together with a central negative dip $W_{cat}\left(0,0\right) =-0.16$. A $p$-SC state is generated when the in-line squeezer works under the condition of amplification with a pump power of $40$ mW, as shown in Figure \ref{fig3}b. Comparing with Figure \ref{fig3}a, the $p$-SC state is squeezed along the $p$ quadrature, and the odd photon-number terms are increased. In contrast, an $x$-SC state is generated when the in-line squeezer works under the condition of deamplification with a pump power of $20$ mW, as shown in Figure \ref{fig3}c. Compared to Figure \ref{fig3}a, the $x$-SC state is squeezed along the $x$ quadrature, and the odd photon-number terms are decreased. All results are corrected for a 80\% detection efficiency, including the transmission efficiency and the homodyne detection efficiency (see Section S1, Supporting Information for more details). Different from the even SC states prepared in previous experiments~\cite{SQDPRL2018,HuangK2015,Ourjoumtsev2007,Etesse2015,Lvovsky2017}, the odd SC states are prepared in our experiment, because the in-line squeezer does not change the parity of an optical cat state.
\begin{figure}[tbp]
\begin{center}
\includegraphics[width=1.0\linewidth]{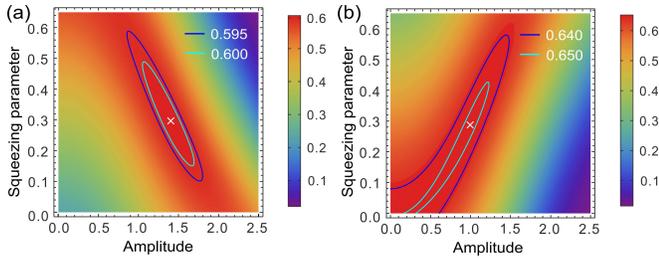}
\end{center}
\caption{The fidelities between the experimentally obtained states and ideal SC states as a function of theoretical amplitude and squeezing parameter. a) Fidelity of the $p$-SC state with $40$ mW pump power for OPA2. b) Fidelity of the $x$-SC state with $20$ mW pump power for OPA2. }
\label{fig4}
\end{figure}

~~The quality of the prepared cat states and SC states is quantified by the fidelity. The fidelity of a cat state is obtained by calculating the similarity between an ideal cat state $\left\vert cat_{-}\right\rangle=\sum\limits_{n=0}^{+\infty }e^{-\left\vert \alpha \right\vert ^{2}/2}\frac{2\alpha ^{2n+1}}{\sqrt{N_{-}(2n+1)!}}$
$\left\vert 2n+1\right\rangle $ and the experimentally reconstructed density matrix $\hat{\rho}^{out}_{c}$, i.e.,
\begin{equation}
F_{cat}=\left\langle cat_{-}\right\vert \hat{\rho}%
^{out}_{c}\left\vert cat_{-}\right\rangle 
\label{eq2}
\end{equation} 
The amplitude and fidelity of the prepared cat state are $\alpha=1.06\pm 0.02$ and $F_{cat}=0.68\pm 0.01$, respectively.

~~The fidelity between the experimentally obtained $\hat{\rho}^{out}_{sc}$
 and an ideal SC state $\left\vert SC_{-}\right\rangle$~is given by 
\begin{equation}
F_{sc}=\left\langle SC\_\right\vert \hat{\rho}^{out}_{sc}\left\vert SC\_\right\rangle 
=\left\langle cat_{-}\right\vert \hat{S}^{\dagger }\left( \zeta \right) \hat{\rho}^{out}_{sc}\hat{S}\left( \zeta \right)\left\vert cat_{-}\right\rangle 
\end{equation}
where  
$\left\vert SC\_\right\rangle =\sum\limits_{n=0}^{+\infty }\hat{S}\left(
\zeta \right) e^{-\left\vert \alpha \right\vert ^{2}/2}\frac{2\alpha ^{2n+1}}{\sqrt{N_{-}(2n+1)!}}\left\vert 2n+1\right\rangle $ represents the odd SC state in the Fock state basis (see Section S2, Supporting Information for more details).  Here, $\hat{S}\left( \zeta \right) =e^{\frac{\zeta ^{\ast }%
\hat{a}^{2}-\zeta \hat{a}^{\dagger 2}}{2}}$ is the
squeezing operator with $\zeta =-re^{i\theta }$; $\hat{a}$ and $\hat{a}^{\dagger}$ are the annihilation and creation operators, respectively. When the $\theta$ is controlled to $0$ or $\pi$, the $p$-SC or $x$-SC state is obtained, respectively. The squeezing parameter and amplitude of SC states are determined by the maximum fidelity. 

~~Figure {\ref{fig4}}a and {\ref{fig4}}b show the fidelities of the $p$-SC state and the $x$-SC state, respectively, in the parameter space of amplitude and the squeezing parameter. The amplitude, squeezing parameter, and fidelity of the $p$-SC state in Figure \ref{fig3}b are $\alpha=1.40\pm 0.03$, $r=0.30\pm 0.02$, and $F_{psc}=0.61\pm 0.01$, respectively. The amplitude, squeezing parameter, and fidelity of the $x$-SC state in Figure \ref{fig3}c are $\alpha=0.99\pm0.01$, $r=0.29\pm 0.01$, and $F_{xsc}=0.65\pm 0.02$, respectively.

~~The generation rates of SC states in previous experiments are limited by the multi-fold photon coincidence measurement or homodyne heralding measurement~\cite{SQDPRL2018,HuangK2015,Ourjoumtsev2007,Etesse2015}. The generation rate of SC states in our experiment is around $2$ kHz, which is  the same as that of the initial cat state, and is higher than that of previous experiments  
\cite{SQDPRL2018,HuangK2015,Ourjoumtsev2007,Etesse2015}. The improvement of the generation rate comes from the fact that only one photon detection is involved and the squeezing operation of the in-line squeezer is deterministic. 

~~The squeezing parameter of the prepared SC states can be actively manipulated by controlling the pump power of the in-line squeezer, enabling the preparation of SC states according to requirements. The experimental results of the prepared SC states at different pump powers of OPA2 are shown in Table 1. The amplitude and the squeezing parameter of $p$-SC states increase with the increase of pump power of OPA2. When manipulating the $x$-SC states, the squeezing parameters are increased whereas the amplitudes are decreased a little bit with the increase of the pump power of OPA2. The fidelities of the $p$-SC and $x$-SC states do not decrease apparently as compared to that of the cat state. The experimental results of the $p$-SC and $x$-SC states apart from those in Figure \ref{fig3} can be found in Section S3, Supporting Information. 

\begin{table}[tbp]
 \caption{Experimental results of SC states.}
\renewcommand\arraystretch{1}

\setlength{\tabcolsep}{1.4mm}{
\begin{tabular}{lllll|||||||||||ccccccc}

\toprule

&\multicolumn{1}{c}{{States}} & \multicolumn{1}{c}{{P}(mW)} & \multicolumn{1}{c}{\ \ $\textit{F}$} & \multicolumn{1}{c}{${\ \ \ \ \ \alpha}$}& \multicolumn{1}{c}{\ \ \ \ \ ${r}$} & \multicolumn{1}{c}{\ \ \ ${W(0,0)}$} \\
\hline

&\multicolumn{1}{c}{} & \multicolumn{1}{c}{$20$} & \multicolumn{1}{c}{\ \ $0.62$  } & \multicolumn{1}{c}{\ \ \ \ \ $1.21$} & \multicolumn{1}{c}{\ \ \ \ \ $0.27$} & \multicolumn{1}{c}{\ \ \ $-0.14$} \\  
&\multicolumn{1}{c}{$p$-SC} & \multicolumn{1}{c}{$30$} & \multicolumn{1}{c}{\ \ $0.62$  } & \multicolumn{1}{c}{\ \ \ \ \ $1.31$} & \multicolumn{1}{c}{\ \ \ \ \  $0.29$} & \multicolumn{1}{c}{\ \ \ $-0.13$} \\ 
&\multicolumn{1}{c}{} & \multicolumn{1}{c}{$40$} & \multicolumn{1}{c}{\ \ $0.61$  } & \multicolumn{1}{c}{\ \ \ \ \ $1.40$} & \multicolumn{1}{c}{\ \ \ \ \ $0.30$} & \multicolumn{1}{c}{\ \ \ $-0.16$} \\ 
\hline

&\multicolumn{1}{c}{} & \multicolumn{1}{c}{$10$} & \multicolumn{1}{c}{\ \ $0.62$  } & \multicolumn{1}{c}{\ \ \ \ \ $1.04$} & \multicolumn{1}{c}{\ \ \ \ \ $0.24$} & \multicolumn{1}{c}{\ \ \ $-0.11$} \\  
&\multicolumn{1}{c}{$x$-SC} & \multicolumn{1}{c}{$15$} & \multicolumn{1}{c}{\ \ $0.65$  } & \multicolumn{1}{c}{\ \ \ \ \ $1.03$} & \multicolumn{1}{c}{\ \ \ \  \ $0.27$} & \multicolumn{1}{c}{\ \ \ $-0.11$}\\ 
&\multicolumn{1}{c}{} & \multicolumn{1}{c}{$20$} & \multicolumn{1}{c}{\ \ $0.65$  } & \multicolumn{1}{c}{\ \ \ \ \ $0.99$} & \multicolumn{1}{c}{\ \ \ \ \ $0.29$} & \multicolumn{1}{c}{\ \ \ $-0.13$}  \\ 
\toprule
\end{tabular}} 
\end{table}

~~In principle, the squeezing operation on the cat state only changes the squeezing parameter of the SC state, while its amplitude keeps unchanged as the cat state. However, it is clear that the amplitudes of the $p$-SC and $x$-SC states increase and decrease in our experiment, respectively. For the $p$-SC state, the overlap between the two component states is decreased, and their quantum interferences are increased. In contrast, for the $x$-SC state, the overlap between the two component states is increased, and their quantum interferences are reduced. There are two possible reasons for the variation of the amplitude. First, the quantum state generated by subtracting one photon from a squeezed vacuum state
$\hat{a}\hat{S}\left(r\right) \left\vert 0\right\rangle$
is only an approximate cat state with amplitude $\alpha$, because the photon number distributions of the two states are similar up to the $3$-photon term, but are different for higher-order terms. If an ideal cat state is squeezed along the superposition direction ($x$ quadrature), $\hat{S}\left(-r\right) \left\vert cat_{-} \right\rangle$, an $x$-SC state with squeezing parameter $r'=r$ and amplitude $\alpha'=\alpha$ will be obtained. While if squeezing photon-subtracted squeezed vacuum state by the same squeezing parameter along the superposition direction, we have $\hat{S}\left(-r\right) \hat{a}\hat{S}\left(r\right) \left\vert0\right\rangle=sinh\left(r\right)\left\vert1\right\rangle$, i.e., a single photon. If fitting it with an $x$-SC state parameterized by an amplitude $\alpha'$ and a squeezing parameter $r'$, we have $r'\rightarrow0$ and an amplitude $\alpha'\rightarrow0$. Thus, the amplitude of the SC state obtained by squeezing the approximate cat state is changed. Second, the antisqueezing level is larger than the squeezing level because of imperfections of the in-line squeezer. The main imperfections come from the intracavity loss, coupling loss of the in-line squeezer and the phase fluctuation~\cite{Takeno2007}. The amplitude of the $p$-SC state is increased, while the amplitude of the $x$-SC state is decreased, when the antisqueezing level is larger than squeezing level of the in-line squeezer. 

\section{Discussion and conclusion} 

In previous experiments, postselection on different quadrature homodyne heralding data is required to manipulate the phase of the quadrature squeezing of an optical SC state \cite{Ourjoumtsev2007,Etesse2015}. Our experiment allows for a much easier manipulation of the phase of the quadrature squeezing, by simply changing the working condition of OPA2. Moreover, conversions between electrical and optical signals are required in the off-line squeezer to realize measurement-based feedback on non-Gaussian states, such as the single photon state and optical cat state~\cite{Miwa2014}. Such conversions inevitably limit the bandwidth of the prepared non-Gaussian states. Instead, our in-line squeezer removes the requirement of electro-optic and opto-electric conversions, allowing for high-bandwidth squeezing operations on non-Gaussian states.

~~In summary, we present a scheme to experimentally prepare an optical SC state and manipulate the phase of the quadrature squeezing via an all-optical in-line squeezer. The phase of the quadrature squeezing and the squeezing parameter
of SC states are manipulated by controlling the working condition and pump power 
of the in-line squeezer. Since the squeezing operation is deterministic, the generation rate of SC states in our experiment is the same as that of the initial cat state. The parity of the prepared SC states is maintained, as the squeezing operation is directly applied on an input cat state. Our results provide a deterministic all-optical measurement-free approach to prepare and manipulate optical SC states without losing the non-classicality, i.e., the negativity of their Wigner functions. This represents a crucial step towards all-optical quantum information processing based on SC states.

\begin{acknowledgments}
This research was supported by the NSFC (Grant No. 11834010, No. 62005149, No. 11974227), Fundamental Research Program of Shanxi Province (Grant No. 20210302121002, No. 20210302122002, No. 201901D211164), Shanxi Scholarship Council of China (No. 2021-03), and the Fund for Shanxi ``1331 Project" Key Subjects Construction.
\end{acknowledgments}

M. W. and M. Z. contributed equally to this work. 

\appendix*
\setcounter{equation}{0}
\setcounter{figure}{0}
\renewcommand\thefigure{A\arabic{figure}}
\renewcommand\theequation{A\arabic{equation}}
\renewcommand\thetable{A\arabic{table}}
\subsection*{Appendix A:~~~Details of experiment}

The SHG cavity is a four-mirror ring cavity consisting of two plane mirrors, two spherical mirrors, and a phase matching periodically poled potassium titanyl phosphate (PPKTP, $1\times 2\times 10$ mm$^{3}$, Raicol) crystal with the nonlinear conversion efficiency 1.8\%/W. The transmissivity of input coupler is $8\%$ at $795$ nm. The SHG cavity is locked by Pound-Drever-Hall (PDH) technique and piezoelectric transducer (PZT) mounted on one plane mirror. OPA1 is also a four-mirror ring cavity with PPKTP ($1\times 2\times 10$ mm$^{3}$) crystal. The transmissivity of output coupler of OPA1 is $12.5\%$ at $795$ nm and the intra-cavity loss is $0.5\%$. A nearly pure squeezed vacuum state with $-3$ dB squeezing is obtained with $25$ mW pump power. OPA2 is also a four-mirror ring cavity with PPKTP ($1\times 2\times 10$ $mm^{3}$) crystal, but  its input coupler with transmissivity of $14.7\%$ at $795$ nm is simultaneously used as output coupler, which is different from OPA1. The intra-cavity loss of OPA2 is $0.4\%$. 

A small fraction ($5\%$) of the produced squeezed vacuum state passes through an interference filter with $0.4$ nm bandwidth and two high finesse ($\sim 1200$) filtering cavities with cavity lengths of $0.75$ mm and $2.05$ mm to filter out the nondegenerate modes of OPA1. Then it is coupled to an Avalanche Photodiode (APD, Excelitas Technologies) through a single-mode fiber. The maximum dark count rate of the APD is $60$ Hz. Each click of the APD heralds the generation of an optical cat state. The residue ($95\%$) of the squeezed vacuum state is detected by the home-made homodyne detector to reconstruct the Wigner function of the cat state. 

The homodyne detection efficiency is limited by the interference efficiency between signal and local oscillator ($98.5\%$), quantum efficiency of photodiodes ($92\%$, S3883, Hamamatsu), and clearance between shot noise and electronic background.  The clearance with $16$ mW local oscillator is more than $21$ dB at $10$ MHz and $18.3$ dB (corresponding to equivalent efficiency of $98.5\%$) at $15$ MHz (bandwidth of OPA2). Thus, the homodyne detection efficiency $\xi $ is their product $98.5\%^{2}\times 92\%\times 98.5\%=87.9\%$. In the experiment, the transmission efficiencies $\eta$ of the optical cat state and squeezed cat (SC) states are both $91.4\%$, where loss mainly comes from an interference filter used to filter the pump beam and other optical elements. Thus, the overall corrected efficiencies (detection efficiencies) of cat state and SC states are both about $\eta \times \xi =80\%$. 

In the experiment, the whole system is operated in the ``sample-and-hold" mode, where $32$ ms for data acquisition and $32$ ms for locking and calibration periods are used periodically. The sequential control is realized by means of two acousto-optic modulators (AOMs) placed on the seed and lock beams of OPA1 and the third AOM before APD.

\begin{figure}[tbp]
\begin{center}
\includegraphics[width=90mm]{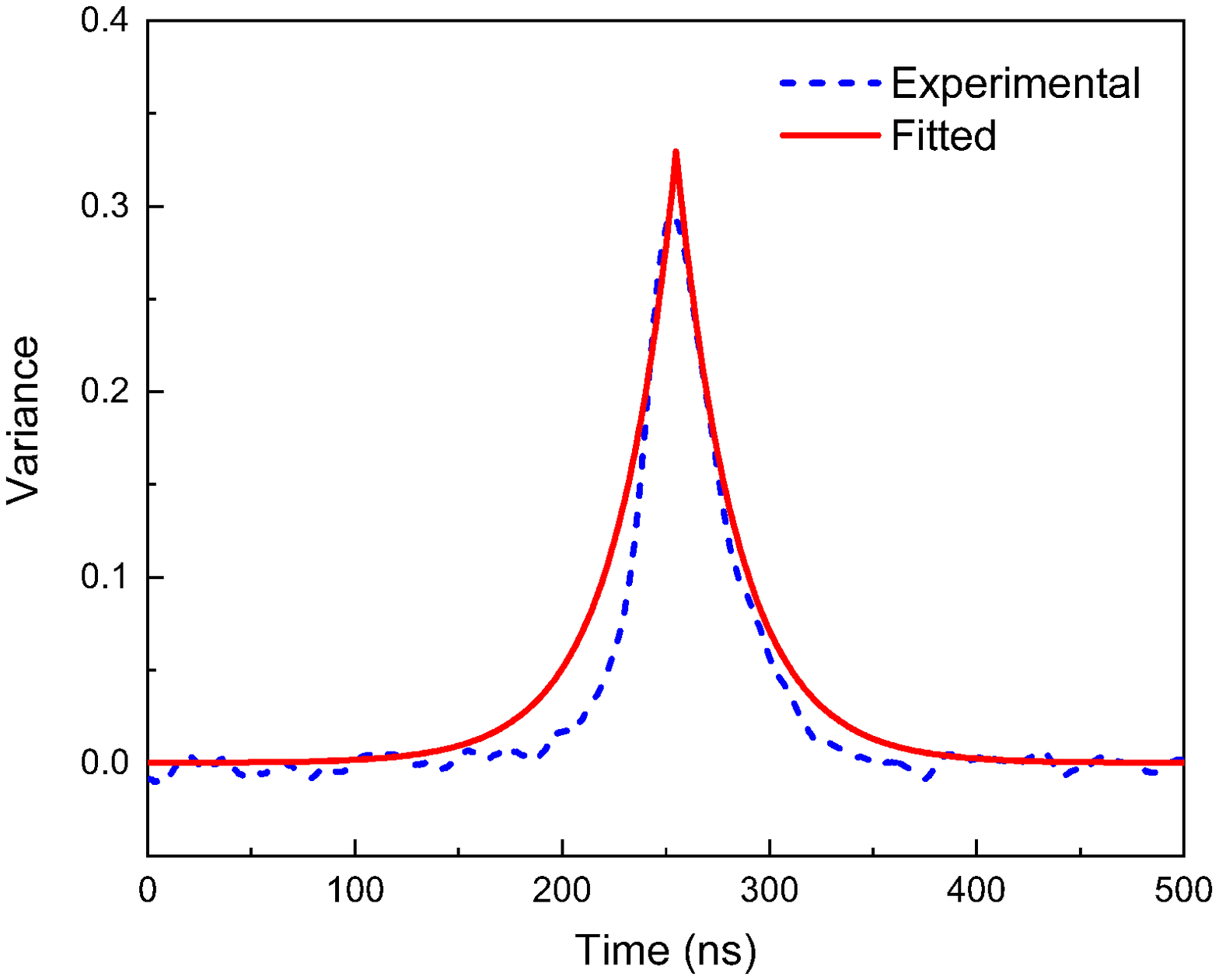}
\end{center}
\caption{
Experimental variance curve and fitted temporal mode of $p$-SC state with $40$ mW pump power of OPA2. Blue dashed curve is  obtained by calculating the variance of $50000$ photocurrent traces. Red solid curve is the fitted temporal mode with the shape of double-side exponential decay.}
\label{figA1}
\end{figure}
\begin{figure*}[tbp]
\begin{center}
\includegraphics[width=115mm]{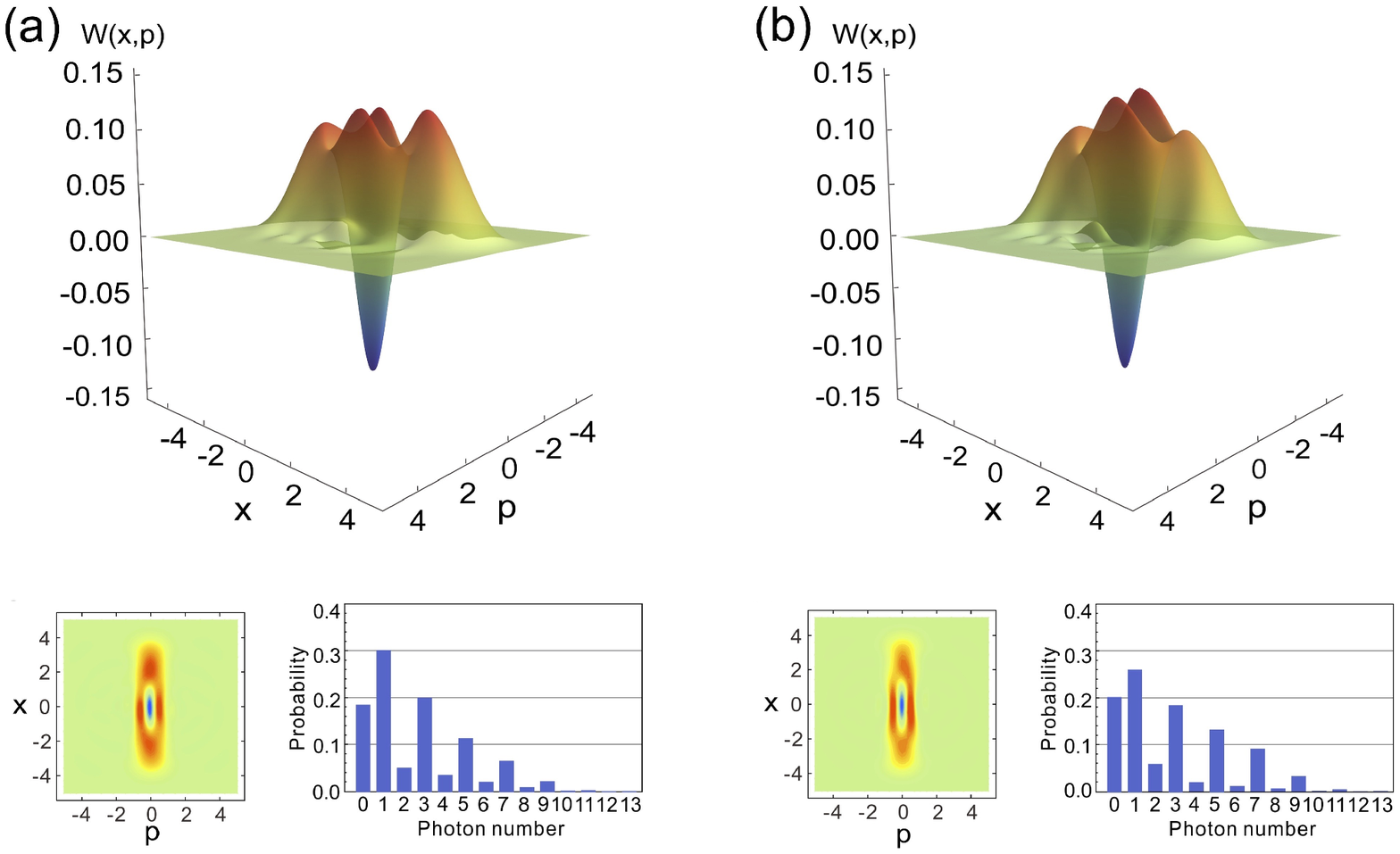}
\end{center}
\caption{Manipulation results of $p$-SC states. (a), (c), (d) The reconstructed Wigner function, the projection and the distribution of photon number of $p$-SC states with $20$ mW pump power of OPA2. (b), (e), (f) The reconstructed Wigner function, the projection and the distribution of photon number of $p$-SC states with $30$ mW pump power of OPA2. The experimental results are corrected for a 80\% detection efficiency.}
\label{figA2}
\end{figure*}
\begin{figure*}[tbp]
\begin{center}
\includegraphics[width=115mm]{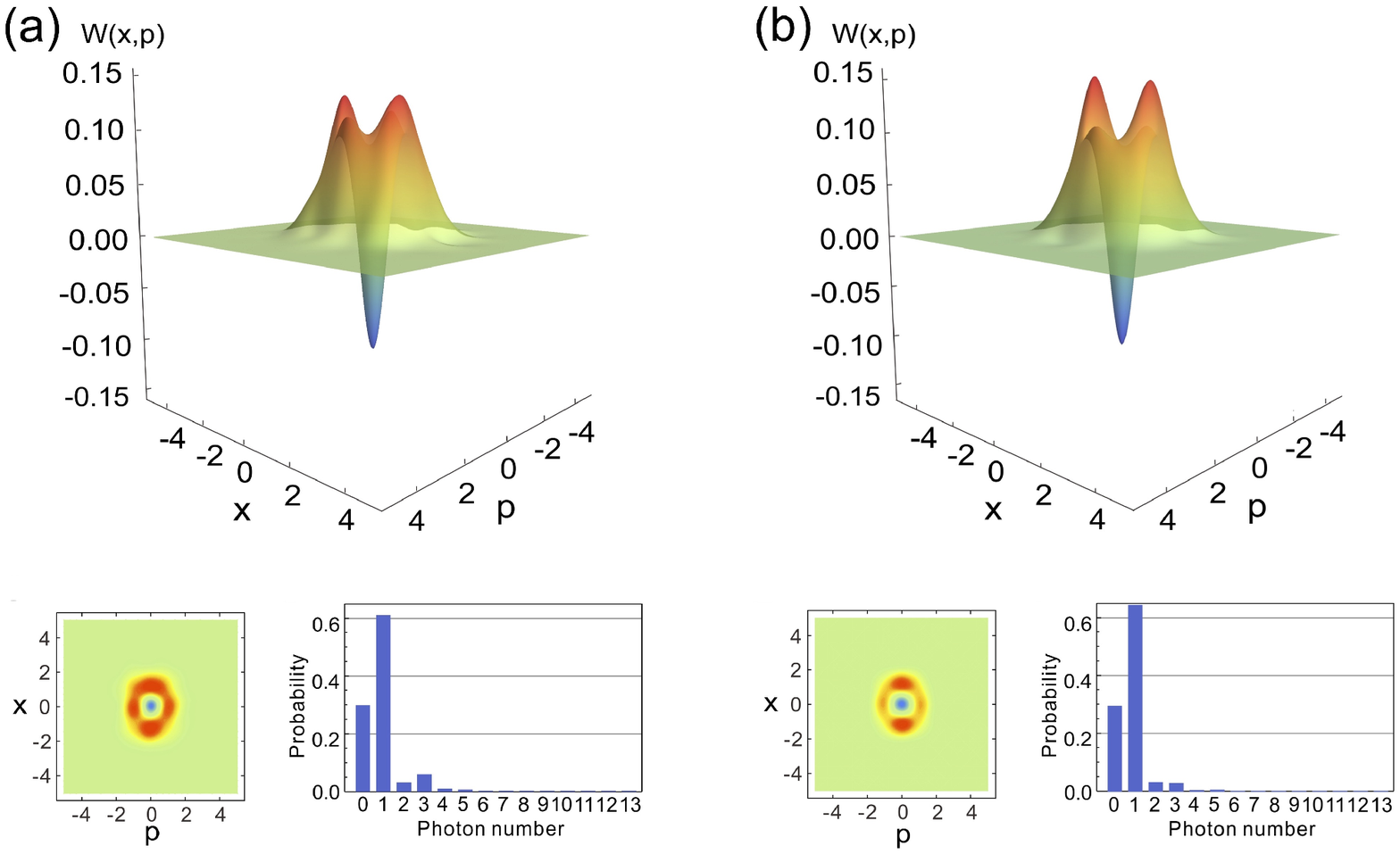}
\end{center}
\caption{Manipulation results of the $x$-SC states. (a), (c), (d) The reconstructed Wigner function, the projection and the distribution of photon number of the $x$-SC state with $10$ mW pump power of OPA2. (b), (e), (f) The reconstructed Wigner function, the projection and the distribution of photon number of the $x$-SC state with $15$ mW pump power of OPA2. The results are corrected for a 80\% detection
efficiency.}
\label{figA3}
\end{figure*}

The output of homodyne detector is divided into direct current (DC) and alternating current (AC). DC part is used for inferring the relative phase between signal and local oscillator by sweeping the PZT, and AC part is used for quadrature measurement. In the ``hold'' mode, the interference fringe from DC part of homodyne detector and voltages for scanning the relative phase are recorded by one oscilloscope, then the dependence of relative phase on the voltage is fitted (cosine function). In the ``sample'' mode, photocurrents of the homodyne detector corresponding to the scanning voltages for every quadrature are recorded by the other broad bandwidth oscilloscope (Lecroy WaveRunner 640 Zi) triggered by clicks of the APD. Thus, the phase for quadrature can be inferred according to the recorded voltage and fitted cosine function. 

In our experiment, $50000$ photocurrent traces of optical cat state are taken and used to reconstruct its Wigner function. Then, $50000$ photocurrent traces of SC states are also taken to reconstruct their Wigner functions at different pump powers. The oscilloscope is triggered by clicks of the APD and works in the segmenting mode with the time of acquisition window $500$ ns for each trigger and with the sample rate $1$ GS/s. So, for a click of the APD, a photocurrent trace $X_\theta(t)$ including 500 Samples is recorded for a segment. The blue dashed curve as shown in Figure \ref{figA1} is obtained by calculating the variance of all photocurrent traces ($50000$ traces) point by point. According to the experimentally obtained variance curve, a smooth temporal mode function $f(t)$ with the shape of double-sided exponential decay (red solid curve as shown in Figure \ref{figA1}) is fitted. In principle, the quadrature value is obtained by integrating the product of homodyne photocurrent and temporal mode in time $X_{\theta}=\int f(t)X_{\theta}(t)dt$. Since the homodyne photocurrents are discrete in time, so the integration is replaced by the sum in time. By multiplying the homodyne photocurrent and fitted temporal mode function point by point, the quadrature value $X_{\theta}=\sum_{1}^{500} f(t)X_{\theta}(t)$ is obtained by summing the $500$ Samples. Repeating this process $50000$ times, we obtain $50000$ quadrature values. 
The relative phase between signal and LO is scanned to make sure that all quadratures are measured. In the process of reconstructing Wigner functions, the iterative maximum-likelihood algorithm \cite{Lvovsky2009} is used.

\subsection*{Appendix B:~~~The fidelities of squeezed cat states}
In the basis of Fock states, an ideal pure odd SC state is expressed as
\begin{equation}
\left\vert SC\_\right\rangle =\sum\limits_{n=0}^{+\infty }\hat{S}\left(
\zeta \right) e^{-\left\vert \alpha \right\vert ^{2}/2}\frac{2\alpha ^{2n+1}}{\sqrt{N_{-}(2n+1)!}}\left\vert 2n+1\right\rangle
\label{EqA1}
\end{equation}
Here, $\hat{S}\left( \zeta \right) =e^{\frac{\zeta ^{\ast }%
\hat{a}^{2}-\zeta \hat{a}^{\dagger 2}}{2}}$ is the
squeezing operator with $\zeta =-re^{i\theta }$, $\hat{a}$ and $\hat{a}^{\dagger}$ are the annihilation and creation operators, respectively.
By substituting the squeezing operator into Eq. \ref{EqA1}, the $p$-SC and $x$-SC states in the Fock state basis are expressed as
\begin{widetext}

\begin{eqnarray}
\left\vert \textit{p-}SC\_\right\rangle 
= \frac{2e^{-\frac{(1+tanhr)\alpha^{2}}{2}}}{\sqrt{N_{-}coshr}}\sum\limits_{n=0}^{+\infty}\frac{{(-\frac{1}{2}tanhr)}^{\frac{2n+1}{2}}}{\sqrt{(2n+1)!}}H_{2n+1}(\frac{-i\alpha}{\sqrt{sinh2r}}) \left\vert 2n+1\right\rangle
\end{eqnarray}
\begin{eqnarray}
\left\vert \textit{x-}SC\_\right\rangle 
= \frac{2e^{\frac{(tanhr-1)\alpha^{2}}{2}}}{\sqrt{N_{-}coshr}}
 \sum\limits_{n=0}^{+\infty}\frac{{(\frac{1}{2}tanhr)}^{\frac{2n+1}{2}}}{\sqrt{(2n+1)!}}H_{2n+1}(\frac{\alpha}{\sqrt{sinh2r}}) \left\vert 2n+1\right\rangle
\end{eqnarray}

\end{widetext}
respectively. The fidelity of experimental prepared $p$-SC ($x$-SC) state can be calculated by substituting experimentally obtained $\hat{\rho}^{out}_{sc}$ and Eq. (A2) [Eq. (A3)] into the Eq. (5) in the main text.

\subsection*{Appendix C:~~~Supplemental experimental results}

Figure {\ref{figA2}} presents the experimentally reconstructed Wigner functions, projections of Wigner functions and photon number distributions of $p$-SC states with different pump powers of OPA2 when the OPA2 works at amplification status. The amplitude, squeezing parameter and fidelity of a $p$-SC state with a pump power of $20$ mW in Figure \ref{figA2} are $\alpha=1.21$, $r=0.27$, and $F=0.62$, respectively. When the pump power of OPA2 is increased to $30$ mW, a $p$-SC state with the amplitude of $\alpha=1.31$, squeezing parameter of $r=0.29$, and fidelity of $F=0.62$ is obtained. 

Figure {\ref{figA3}} presents the experimentally reconstructed Wigner functions, projections of Wigner functions and photon number distributions of $x$-SC states with $10$ mW and $15$ mW pump powers of OPA2. The amplitude, squeezing parameter and fidelity of the $x$-SC state with a pump power of $10$ mW in Figure \ref{figA3} are $\alpha=1.04$, $r=0.24$, and $F=0.62$, respectively. When the pump power of OPA2 is increased to $15$ mW, the $x$-SC state with the amplitude of $\alpha=1.03$, squeezing parameter of $r=0.27$, and fidelity of $F=0.65$ is obtained.

\end{document}